\documentclass[aps,twocolumn,prl,showpacs,amsmath,amssymb,10pt]{revtex4-2}
\usepackage{graphicx}
\usepackage{xcolor}

\usepackage{float}
\usepackage{scrextend}
\usepackage{manfnt,pifont}
\usepackage{hyperref}
\usepackage{url}

\usepackage[normalem]{ulem}

\usepackage[outline]{contour}
\usepackage{textgreek}

\usepackage{tikz}
\usetikzlibrary{shapes}

\usepackage{CJK}
\usepackage{enumitem}
\setlist[itemize]{leftmargin=*}

\newcommand\wh\widehat

\newcommand\ri{{\mathrm{i}}}

\newcommand\rd{{\mathrm{d}}}

\makeatletter
\DeclareFontFamily{OMX}{MnSymbolE}{}
\DeclareSymbolFont{MnLargeSymbols}{OMX}{MnSymbolE}{m}{n}
\SetSymbolFont{MnLargeSymbols}{bold}{OMX}{MnSymbolE}{b}{n}
\DeclareFontShape{OMX}{MnSymbolE}{m}{n}{
    <-6>  MnSymbolE5
   <6-7>  MnSymbolE6
   <7-8>  MnSymbolE7
   <8-9>  MnSymbolE8
   <9-10> MnSymbolE9
  <10-12> MnSymbolE10
  <12->   MnSymbolE-Bold12
}{}
\DeclareFontShape{OMX}{MnSymbolE}{b}{n}{
    <-6>  MnSymbolE-Bold5
   <6-7>  MnSymbolE-Bold6
   <7-8>  MnSymbolE-Bold7
   <8-9>  MnSymbolE-Bold8
   <9-10> MnSymbolE-Bold9
  <10-12> MnSymbolE-Bold10
  <12->   MnSymbolE12
}{}

\let\llangle\@undefined
\let\rrangle\@undefined
\DeclareMathDelimiter{\llangle}{\mathopen}
                     {MnLargeSymbols}{'164}{MnLargeSymbols}{'164}
\DeclareMathDelimiter{\rrangle}{\mathclose}
                     {MnLargeSymbols}{'171}{MnLargeSymbols}{'171}
\makeatother

\usepackage{eso-pic}

\begin{document}

%\reportnum{-3}{USTC-ICTS/PCFT-26-30}

\title{A universal scaling function for giant graviton OPE coefficients}

\author{Miao He$^{a}$}

\author{Yunfeng Jiang$^{b,c}$}
\thanks{Corresponding author: jinagyf2008@seu.edu.cn}

%\author{Yunfeng Jiang$^{a,c}$}

%\author{Yang Zhang$^{b,c,d}$}
%\thanks{Corresponding author: yzhphy@ustc.edu.cn}

\affiliation{$^a$School of Physics and Electronic Information, Anhui Normal University, Wuhu, Anhui 241000, China}
\affiliation{$^b$School of Physics \& Shing-Tung Yau Center, Southeast University, Nanjing 211189, P. R. China}
\affiliation{$^c$Peng Huanwu Center for Fundamental Theory, Hefei, Anhui 230026, China}

\begin{abstract}
We consider the large spin limit $S\to\infty$ of the OPE coefficient for two maximal giant gravitons and a spinning operator with finite twist in planar $\mathcal{N}=4$ super-Yang-Mills theory. Up to an overall normalization factor, this OPE coefficient exhibits a power-law scaling of the form $S^{d(g)}$. We provide strong evidence that the exponent $d(g)$ is free from finite size corrections. Consequently, the all-loop asymptotic expression for the OPE coefficient can be used to determine $d(g)$ exactly. We develop a systematic method to compute the scaling function $d(g)$ for arbitrary values of the 't Hooft coupling $g$. At weak coupling, our result agrees perfectly with the available field-theoretic results up to three-loop order. At strong coupling, we present the first three orders as concrete predictions. We further provide the finite-coupling result and show that it interpolates smoothly between the weak- and strong-coupling regimes.
\end{abstract}

\maketitle

\noindent{\bf Introduction.}  
The AdS/CFT correspondence encapsulates a profound conceptual insight: a conformal field theory and a gravitational theory in anti-de Sitter space are two sides of the same coin. Although these two theoretical frameworks appear markedly different, they in fact describe identical physics. Weakly coupled regimes are naturally interpreted in terms of quantum field theory, while strongly coupled regimes admit a gravitational description. This duality, however, poses significant challenges for explicit verification, particularly for non-BPS observables. Ideally, one would like to compute a given quantity across the entire coupling range: recovering the quantum field theory result at weak coupling and reproducing the gravity or string theory prediction at strong coupling. Achieving this is inherently difficult, as it requires a method valid at finite coupling.

A landmark example where this program has been successfully realized is the duality between $\mathcal{N}=4$ super-Yang–Mills ($\mathcal{N}=4$ SYM) theory and Type-IIB superstring theory on AdS$_5\times S^5$. The quantity in question is the cusp anomalous dimension. At weak coupling, it is related to the vacuum expectation value of a light-like Wilson loop with a cusp and can be computed by field theoretic methods \cite{Kotikov:2004er,Bern:2006ew,Henn:2019swt}, while at strong coupling it corresponds to the energy density of the GKP string \cite{Gubser:2002tv,Frolov:2002av,Kruczenski:2002fb,Kruczenski:2007cy,Roiban:2007jf}. These two limits can thus be addressed using entirely different computational techniques. Remarkably, integrability allows for an exact determination of this quantity at arbitrary coupling \cite{Eden:2006rx,Beisert:2006ib,Beisert:2006ez,Benna:2006nd,Basso:2007wd}. The resulting interpolation between weak and strong coupling is smooth and matches perturbative expansions with striking precision at both ends, thereby providing an extraordinarily nontrivial test of the AdS/CFT. 

Within the integrability framework, the cusp anomalous dimension is accessed via the anomalous dimension of the twist-two spinning operator 
$\mathcal{O}_S=\text{Tr}(ZD^SZ)+\ldots$, where $Z$ denotes a complex scalar field, $D$ is a covariant derivative along a light-like direction and $S$ is the Lorentz spin. In the large spin limit $S\to\infty$, the anomalous dimension exhibits a logarithmic scaling of the form $\gamma_S(g)=f(g)\log S+\ldots$, where the coefficient $f(g)$ is (twice) the cusp anomalous dimension. Advances in integrability-based methods for solving the spectral problem \cite{Beisert:2005fw,Gromov:2009tv,Arutyunov:2009ur,Gromov:2013pga} have enabled highly non-trivial tests of AdS/CFT correspondence at the spectral level.

Nevertheless, analogous precision tests at the level of operator product expansion (OPE) coefficients remain absent to date. Despite substantial progress in computing OPE coefficients \cite{Escobedo:2010xs,Foda:2011rr,Gromov:2012vu,Caetano:2016keh,Vieira:2013wya,Jiang:2014mja,Basso:2015zoa,Basso:2015eqa,Basso:2017muf,Jiang:2015lda,Bercini:2022jxo,Bargheer:2026kon,Basso:2022nny,Jiang:2019xdz,Jiang:2019zig,Janik:2010gc,Zarembo:2010rr,Costa:2010rz,Roiban:2010fe,Bajnok:2014sza,Janik:2011bd,Kazama:2011cp,Kazama:2012is,Kazama:2013qsa}, an exact finite-coupling result is still out of reach, mainly due to the difficulty in accounting for all finite size corrections in a systematic manner. Notably, in the spectral problem, the cusp anomalous dimension is free from such corrections, which allowed for its exact determination at arbitrary coupling even prior to the development of finite-size techniques. It would therefore be of great interest to identify an OPE coefficient analogue that shares this property \emph{i.e.}, one that does not receive finite-size corrections and can be computed exactly at finite coupling using asymptotic results. Such a quantity would furnish a solid foundation for an exact test of the AdS/CFT correspondence at the level of OPE coefficients. The present work aims to propose such a quantity and to present its explicit computation at all values of the coupling.

More precisely, we propose that the large spin limit of the OPE coefficient involving two maximal giant gravitons and one twist-two operator serves as such a quantity. Denoting this OPE coefficient by $\mathfrak{D}_S(g)$, we find that in the large spin limit it exhibits the scaling behavior $\mathfrak{D}_S(g)=2^{-S+\frac{1}{2}} S^{d(g)}+\ldots$, where the ellipsis denotes subleading terms. We present strong evidence that the exponent $d(g)$ is free from finite-size corrections, thereby rendering it accessible via the asymptotic results derived in \cite{Jiang:2019xdz,Jiang:2019zig}. We develop an integrability-based method that computes $d(g)$ for arbitrary values of the coupling constant $g$.

At weak coupling, we perform a perturbative expansion and obtain perfect agreement with existing results up to three-loop order. In the strong-coupling regime, where no string-theory prediction is currently available, we provide concrete analytic predictions for the first two leading orders in the large-$g$ expansion. Furthermore, we compute the coefficient at finite coupling numerically and demonstrate a smooth interpolation between the weak- and strong-coupling limits.

\vspace{0.5cm}

\noindent{\bf Giant graviton OPE coefficient.}
The OPE coefficient of interest is encoded in the three-point function involving two determinant operators $\mathcal{D}_{1,2}\equiv\det\mathcal{Z}(a_{1,2})$ and a single-trace twist two operator $\mathcal{O}(a_3)=\text{Tr}(\mathcal{Z}D^S\mathcal{Z})(a_3)+\cdots$ with $\mathcal{Z}(a)$ is a complex scalar field defined as \cite{Drukker:2009sf}
\begin{align}
\mathcal{Z}(a)\equiv\frac{(1+a^2)\Phi^1+\ri(1-a^2)\Phi^2+2\ri a\Phi^4}{\sqrt{2}},
\end{align}
evaluated at the spacetime point $x^{\mu}=(0,a,0,0)$. Here $\Phi^{1,2,4}$ are real scalar fields of $\mathcal{N}=4$ SYM theory. Assuming the twist-two operator is unit-normalized, the spacetime dependence of the three-point function is fixed up to an overall constant
\begin{align}
\langle\mathcal{D}_1\mathcal{D}_2\mathcal{O}_S(0)\rangle=\left(\frac{a_1-a_2}{a_1a_2}\right)^{\Delta_S(g)-2}\mathfrak{D}_S(g),
\end{align}
where $\Delta_S(g)$ is the scaling dimension of the twist-2 operator and $\mathfrak{D}_S(g)$ denotes the OPE coefficient that we aim to study. In this work, the 't Hooft coupling defined as $g^2=g_{\text{YM}}^2N_c/16\pi^2$.

\vspace{0.3cm}
\noindent{\it - Asymptotic result.}  In \cite{Jiang:2019xdz,Jiang:2019zig}, an integrability-based method was developed for computing $\mathfrak{D}_S(g)$. This approach reformulates the OPE coefficient as a worldsheet $g$-function, in which all finite-size corrections can be systematically accounted for via the thermodynamic Bethe ansatz. Although the method is in principle exact, its full implementation requires evaluating an infinite-dimensional Fredholm determinant that depends on infinitely many $Y$-functions. Consequently, turning this formalism into a practical and efficient computational tool remains a further work, and an explicit calculation of the OPE coefficient at finite $g$ is not yet available. Nevertheless, with this non-perturbative framework, one can derive an asymptotic result that is valid before finite-size effects become relevant, and is exact for non-BPS operators with large $R$-charge. Our analysis begins with the exact asymptotic expression
\begin{align}
\label{eq:asympD}
\overline{\mathfrak{D}}_S(g)=2^{-S+\frac{1}{2}}\sqrt{\prod_{j=1}^{S/2}\frac{u_j^2+\frac{1}{4}}{u_j^2}\sigma^2_B(u_j)}\times\sqrt{\frac{\det G_+}{\det G_-}},
\end{align}
where the overbar indicates that the result is asymptotic.  Here $\sigma_B(u)$ is the boundary dressing phase derived in~\cite{Jiang:2019xdz} and the set $\{u_j\}$ is the solution of the following asymptotic Bethe ansatz equations (BAEs)
\begin{align}
\label{eq:asymBAE}
\left( \frac{x_{k}^{+}} {x_{k}^{-}} \right)^{L}&=\prod_{\substack{j=1\\j \neq k}}^{S} \frac{x_{k}^{-}-x_{j}^{+}} {x_{k}^{+}-x_{j}^{-}} \, \frac{1-1/x_{k}^{+} x_{j}^{-}} {1-1/x_{k}^{-} x_{j}^{+}}e^{2\ri\theta(u_k,u_j)},
\end{align}
with $L=2$. In the above, $x_k^\pm\equiv x(u_k\pm\frac{\ri}{2})$ and $x(u)=(u+\sqrt{u^2-4g^2})/2g$ is the Zhukowsky variable, $\theta(u,v)$ is the BES dressing phase~\cite{Beisert:2006ez}. Finally, the Gaudin-like determinants $\det G_{\pm}$ in~\eqref{eq:asympD} are associated with the BAEs. Their explicit form will be provided in the supplementary material.

\vspace{0.3cm}
\noindent{\it - Large spin limit.} The OPE coefficient $\mathfrak{D}_S(g)$ can be extracted from the four-point function involving two giant gravitons and two $\frac{1}{2}$-BPS $\mathbf{20}'$, as detailed in~\cite{Wu:2025ott,Jiang:2023uut,He:2026ios}. Explicit results at weak coupling show that in the large spin limit $S\to\infty$, the OPE coefficient exhibits the following simple scaling behavior
\begin{align}
\lim_{S\to\infty}\log\left(\mathfrak{D}_S(g)\right)=-(\log2)S+d(g)\log S+\ldots
\end{align}
where $d(g)$ can be extracted from the OPE coefficient. We will demonstrate this scaling behavior from the integrability result. It is worth noting that the OPE coefficients involving non-maximal giant gravitons do not display such a simple logarithmic scaling~\cite{He:2026ios}.

We will confirm this scaling behavior from the integrability side by taking the large spin limit of the asymptotic expression \eqref{eq:asympD}. We then provide evidence that the coefficient $d(g)$ is free from finite-size corrections. This allows us to extend our analysis to all-loops and extract $d(g)$ non-perturbatively.

\vspace{0.3cm}
\noindent{\it - Leading order.} In the large spin limit, the number of magnons $S$ diverges, and it is convenient to describe the Bethe roots in terms of a continuous density. Following the treatment of~\cite{Eden:2006rx}, we introduce the rescaled variables $\bar{u}_k=u_k/S$, so that the roots are distributed over the interval $[-\frac{1}{2},\frac{1}{2}]$, and sums over Bethe roots are replaced by integrals. The leading-order density function was derived in~\cite{Eden:2006rx} and reads
\begin{align}
\label{eq:leadingDensity}
\rho_0(\bar{u})=\frac{1}{\pi}\log\frac{1+\sqrt{1-4\bar{u}^2}}{1-\sqrt{1-4\bar{u}^2}}\,.
\end{align}
Taking the logarithm of the asymptotic result \eqref{eq:asympD}, and omitting the trivial factor $2^{-S+\frac{1}{2}}$, the remaining contribution separates naturally into two parts: the prefactor and the ratio of Gaudin-like determinants, which we denote by $\mathcal{P}_0(S)$ and $\mathcal{R}_0(S)$ respectively. The prefactor is given by
\begin{align}
\mathcal{P}_0(S)=
\frac{1}{4}\sum_{j=1}^S\log\left(\frac{u_j^2+\frac{1}{4}}{u_j^2}\right)\,.
\end{align}
In the large-$S$ limit, the prefactor can be written as an integral in terms of the rescaled variable, which reads 
\begin{align}
\mathcal{P}_0(S)=&\frac{S}{4}\int_{-1/2}^{1/2}\log\left(\frac{\bar{u}^2+\frac{1}{4S^2}}{\bar{u}^2} \right)\rho_0(\bar{u})\rd \bar{u}\\\nonumber
=&\frac{1}{2}\log S+\mathcal{O}(S^0).
\end{align}
We now turn to the ratio of Gaudin-like determinants
\begin{align}
\mathcal{R}_0(S)=\frac{1}{2}\log\left(\frac{\det G_+}{\det G_-}\right).
\end{align}
To facilitate the large-spin analysis, we express the Gaudin-like determinants as
\begin{align}
\det G_{\pm}=\prod_{j=1}^{S/2}m_j\,\times\det(1- K_{\pm}),
\end{align}
where
\begin{align}
(K_\pm)_{jk}=\frac{1}{m_j}\left(\varphi(u_j,u_k)\pm\varphi(u_j,-u_k)\right),
\end{align}
with
\begin{align}
m_j&=-\frac{L}{u_j^2+\frac{1}{4}}-\sum_{k=1}^S\frac{2}{(u_j-u_k)^2+1},\\
\varphi(u,v)&=-\frac{2}{(u-v)^2+1}\,.
\end{align}
We can then rewrite the ratio of determinants as
\begin{align}
\mathcal{R}_0(S)=&\sum_{n=1}^\infty\frac{(\text{Tr}(K^+)^n-\text{Tr}(K^-)^n)}{2n}\,.
\end{align}
In the large spin limit, the diagonal elements $m_j$ for the ground state scales as
\begin{align}
\label{eq:mjlargeS}
\lim_{S\to\infty} m_j=2\pi\rho_0(\bar{u}_j),
\end{align}
which can be obtained by taking the derivative of logarithm of BAEs. The trace in the same limit can be written in terms of the multiple integrals against the density. The density $\rho_0(\bar{u}_j)$ appearing in the numerator then cancels precisely with the factor of $m_j$ from \eqref{eq:mjlargeS}, leading to a significant simplification. Since we consider the parity invariant configuration of Bethe roots, the ratio of determinants can ultimately be written as
\begin{align}
\label{eq:trace_n_1}
\mathcal{R}_0(S)=-\sum_{n=1}^{\infty}\frac{1}{2n}\int_{-S/2}^{S/2}\prod_{j=1}^{n}\frac{du_j}{2\pi}\varphi(u_{j},u_{j+1}), 
\end{align}    
where $u_{n+1}=-u_1$. Carrying out the resulting integrals in large $S$ limit yields
\begin{align}
\label{eq:tree_R}
\mathcal{R}_0(S)=-\sum_{n=1}^{\infty}\frac{1}{2n\pi}\arctan\left(\frac{S}{n}\right)=-\frac{1}{4}\log S+\mathcal{O}(S^0).
\end{align}
Combining everything, we find that at leading order
\begin{align}
\log\mathfrak{D}_S(0)=-(\log 2)S+\frac{1}{4}\log S+\ldots
\end{align}

\vspace{0.3cm}
\noindent{\it - Universal scaling function.} We now turn to the argument that the scaling function $d(g)$ receives no finite-size corrections. From the leading-order analysis, it is evident that in the scaling limit there are two potential sources of $L$-dependence: the asymptotic Bethe ansatz equations, which depend explicitly on $L$, and the ratio of Gaudin-like determinants $\mathcal{R}_0(S)$. Concerning the BAEs, it is already known from the computation of the cusp anomalous dimension that, for the lowest-energy state in the large-spin limit and with $L$ held finite, the resulting density $\rho(\bar{u})$ is universal and independent of $L$. Consequently, once the density is universal, the prefactor, being obtained by integrating a known function against this density, also becomes $L$-independent. This is precisely the same reasoning that underlies the absence of finite-size corrections for the cusp anomalous dimension.

The other possible source of finite-size effects resides within the Gaudin-like determinants, specifically in the quantities $m_j$. However, as we have demonstrated in the large spin limit, the terms proportional to $L$ drop out, and the limiting expression is determined solely by the density. Hence, the ratio of Gaudin-like determinants is likewise independent of the twist $L$. 
More specifically, taking the derivative with respect to $u_k$ on both sides of logarithm of the asymptotic BAEs~\eqref{eq:asymBAE}, the all-loop quantities $M_j$ entering the Gaudin-like determinants are proportional to the derivative of the counting function evaluated at the corresponding Bethe roots, which becomes the density in large spin limit. For the ground state, one may identify
\begin{align}
\lim_{S\to\infty}M_j=2\pi\lim_{S\to\infty}\frac{\rd \tilde{n}_k}{\rd u_k}=2\pi\rho(u).
\end{align} 
This relation is verified numerically up to three-loop in Figure~\ref{total_density_m} and more details can be found in supplemental material. It is worth noting that the ratio of determinants reduces to a series of multiple integrals involving only the scattering kernel. It is therefore independent of the twist $L$ and contains no information about the Bethe roots.
It follows from our derivation of the large spin limit that all $L$-dependence cancels for the ground state, rendering the scaling function universal for any finite value of $L$.

\begin{figure} 
    \centering           \includegraphics[width=0.47\textwidth]{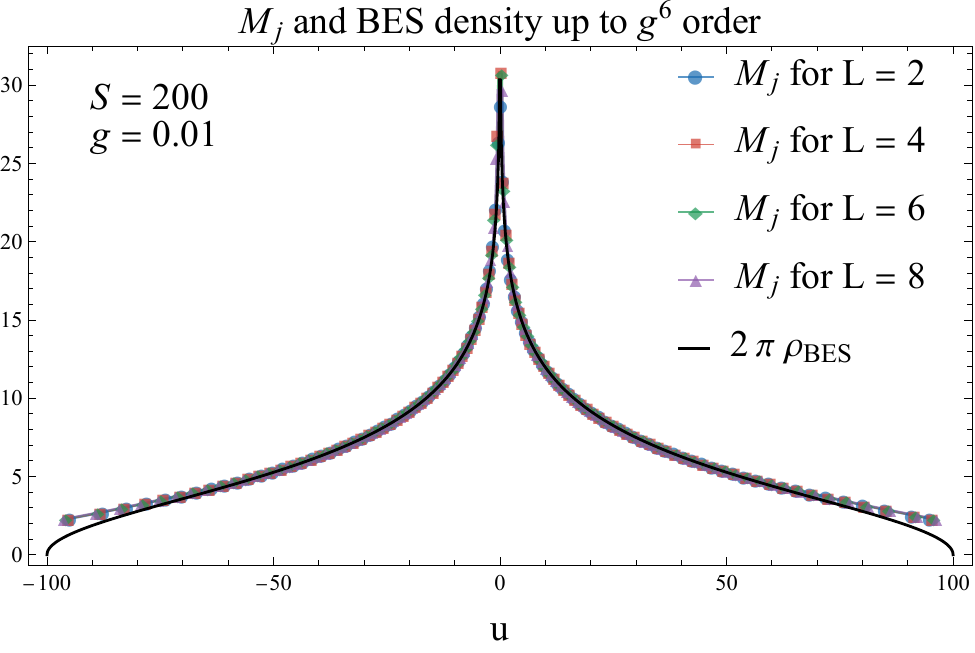}
    \caption{We plot the $M_j$ up to $g^6$ order for different twists $L=2,4,6,8$ with $S=200$. The result suggests that $M_j$ is twist-independent and tends to the BES density in large spin limit. }
    \label{total_density_m} 
\end{figure}

We emphasize that the above argument should be regarded as strong evidence for the absence of finite-size corrections in the scaling function, rather than as a rigorous proof. The reasoning parallels that used for the cusp anomalous dimension, though it would certainly be desirable to subject it to further independent tests. In this work, we adopt this statement as a conjecture, motivated by the consistency of the argument. At low orders in the weak coupling limit where explicit results are available for comparison, we have verified that this conjecture holds true.

\vspace{0.5cm}

\noindent{\bf All-loop result.} Assuming that the scaling function is indeed free from finite-size corrections, we now develop a method to compute it at arbitrary coupling by taking the large spin limit of the asymptotic result.

\vspace{0.3cm}
\noindent{\it - BES equation.} At finite coupling, the asymptotic Bethe ansatz equations in the large spin limit can be reformulated as an integral equation known as the BES equation \cite{Beisert:2006ez}. Following the approach of \cite{Eden:2006rx,Beisert:2006ez}, we decompose the density of Bethe roots into a one-loop piece $\rho_0(u)$ and higher order piece $\tilde{\sigma}(u)$, such that $\rho(u)=\rho_0(u)+\tilde{\sigma}(u)$ where $\rho_0(u)$ is given in \eqref{eq:leadingDensity} up to a variable rescaling. The scale of the density fluctuation allows us to define $\tilde{\sigma}(u)=-2g^2E_0\sigma(u)/S$, where $E_0$ denotes the energy of one-loop ground state. The Fourier-type transformed density fluctuation is known to satisfy the BES equation~\cite{Beisert:2006ez}. The BES equation can be solved either analytically in the weak and strong coupling limits, or numerically for arbitrary finite values of the coupling $g$.

\vspace{0.3cm}
\noindent{\it - Prefactor.} In the large spin limit, the all-loop prefactor
\begin{align}
\mathcal{P}(S)=\frac{1}{4}\sum_{j=1}^S\left[\log\left(\frac{u_j^2+\frac{1}{4}}{u_j^2}\right)+\log\sigma_B^2(u_j)\right],
\end{align}
can be expressed as
\begin{align}
\mathcal{P}(S)=&\,\mathcal{P}_0(S)-\frac{g^2E_0}{2}\int_{-\infty}^{\infty}\log\left(\frac{u^2+\frac{1}{4}}{u^2}\right)\sigma(u)\rd u\\\nonumber
&+\frac{S}{4}\int_{-\infty}^{\infty}\log\sigma_B^2(u)\,\rho(u)\rd u\,.
\end{align}
where $\mathcal{P}_0(S)$ is the leading order contribution defined earlier. Upon substituting the explicit forms of the densities and the boundary dressing phase in the large spin limit, $\mathcal{P}(S)$ can be evaluated straightforwardly.

\vspace{0.3cm}
\noindent{\it - Ratio of determinants.} We now consider the large spin limit of the all-loop ratio of determinants. Although the contribution from the ratio of determinants may appear rather involved, we find the surprising result that all higher-loop corrections vanish in the large-spin limit, namely
\begin{align}
\lim_{S\to\infty}\mathcal{R}(S)=\lim_{S\to \infty}\mathcal{R}_0(S)=-\frac{1}{4}\log S\,.
\end{align}
We briefly explain the reasoning here and relegate the details to the supplementary material. First, as in the leading-order derivation, the density of states cancels out in the large spin limit, so that $\mathcal{R}(S)$ becomes independent of $\rho(u)$ and reduces to a universal function of the kernel alone, which greatly simplifying its evaluation. Second, a careful analysis shows that higher-order corrections in the large-$S$ expansion start only at order $1/S$, and therefore do not contribute to the $\log S$ term that determines the scaling function. This observation not only tremendously simplifies our computation, but also highlights the relative simplicity of the scaling function $d(g)$ as compared to the full OPE coefficient.

\vspace{0.5cm}
\noindent{\bf Results for the scaling function.} 
The upshot of the preceding section is that extracting the scaling function $d(g)$ requires only the large spin limit of $\mathcal{P}(S)$, from which we read off the coefficient of $\log S$. Following this strategy, we now present explicit results for the scaling function at weak, strong, and finite coupling.

\vspace{0.3cm}
\noindent{\it - Weak coupling.} 
The weak-coupling expansion of the density fluctuation can be obtained using the method developed in~\cite{Eden:2006rx, Beisert:2006ez}, where the first few orders are also given. Likewise, the weak-coupling expansion of the boundary dressing phase is available from~\cite{Jiang:2019xdz}. Substituting these ingredients, we obtain the weak-coupling series for $d(g)$. The result up to four-loop order reads
\begin{align}
&d(g)|_{0<g\ll 1}=\frac{1}{4}-(4\log2)\,g^2\\\nonumber
&+\frac{4}{3} \Big(\pi ^2 \log 2-9 \zeta (3)\Big)g^4\\\nonumber
&+\frac{4}{45}\Big(-11 \pi ^4 \log (2)+30 \pi ^2 \zeta (3)+1350 \zeta (5)\Big)g^6\\\nonumber
&+\frac{4}{315} \Big(-119 \pi ^4 \zeta (3)-2100 \pi ^2 \zeta (5)+73 \pi ^6\log 2\\\nonumber
&\quad +315 \left(8 \zeta (3)^2 \log 2-315 \zeta (7)\right)\Big)g^8+\mathcal{O}(g^{10}).
\end{align}
This expression agrees perfectly with the recent field-theory computation reported in~\cite{He:2026ios} up to three-loop. The four-loop result can be viewed as a prediction.

\vspace{0.3cm}
\noindent{\it - Strong coupling.} 
At strong coupling, we employ the method of~\cite{Basso:2007wd} to determine the density of Bethe roots, and the boundary dressing phase can likewise be expanded in the strong-coupling regime. Substituting these expressions yields
\begin{align}
d(g)|_{g\gg 1}=&\,-(4\log 2) g+\left(\frac{3\log^2 2}{\pi}+\frac{2\sqrt{2}-1}{4}\right)\\\nonumber
&+\left(\frac{1}{16\sqrt{2}}+\frac{K\log 2}{4\pi^2}\right)\frac{1}{g}+\mathcal{O}(g^{-2}),
\end{align}
where $K=0.91596\cdots$ is the Catalan constant. So far no string-theoretic computation exists for the OPE coefficient of two giant gravitons and one twist-two operator at strong coupling in the large spin limit. The only available results involve two giant gravitons with one or two BPS operators in the supergravity limit \cite{Yang:2021kot,Chen:2025yxg,Chen:2026ium,Brown:2024tru,Brown:2025huy}. Extracting the OPE coefficient for large spin twist-two operators would require a computation involving the GKP string. Our result thus provides a concrete prediction for future strong-coupling checks.

\begin{figure} 
    \centering       
    \includegraphics[width=0.48\textwidth]{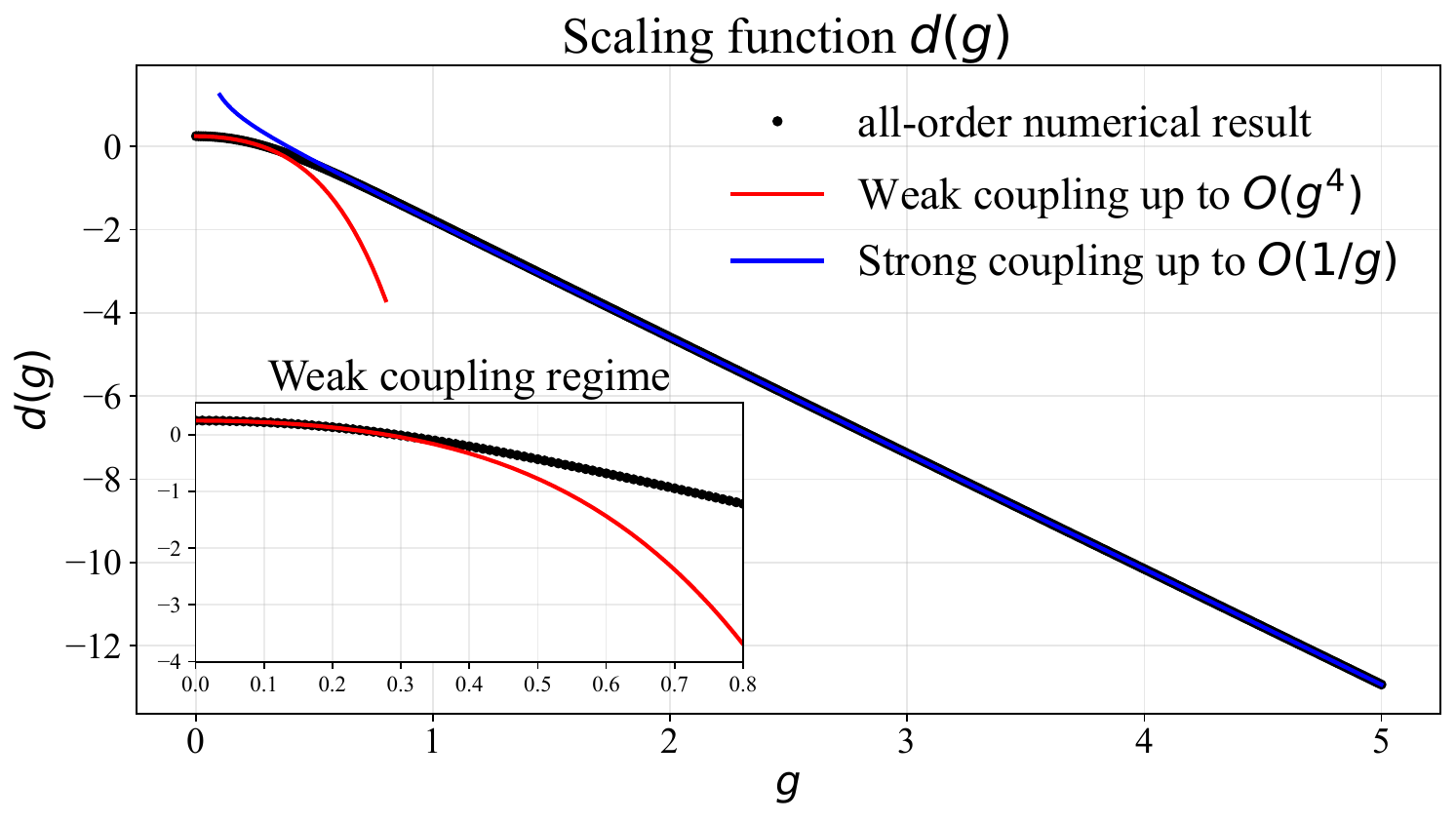}
    \caption{The scaling function is shown. The black dots represent the all order numerical result. The red line is the weak coupling expansion and the blue line is the strong coupling expansion. The numerical results are accurately reproduced by both the weak- and strong-coupling expansions.}
    \label{all_order_scaling_function} 
\end{figure}

\vspace{0.3cm}
\noindent{\it - Finite coupling.} For finite values of the coupling, we adapt the Bessel-matrix method developed in~\cite{Benna:2006nd} to the numerical evaluation of the boundary dressing phase. With this approach, we obtain the scaling function $d(g)$ for arbitrary finite $g$. The resulting curve is displayed in Fig.~\ref{all_order_scaling_function}, where we also observe a smooth interpolation between the weak- and strong-coupling regimes.

\vspace{0.5cm}
\noindent{\bf Discussions.} In this work, we have investigated the OPE coefficient involving two giant gravitons and one twist-two operator. In the large spin limit $S\to\infty$, and up to an overall normalization factor, this OPE coefficient exhibits a power-law behavior of the form $S^{d(g)}$, where the exponent $d(g)$ is a function of the coupling. We have provided strong evidence that this scaling function is free from finite-size corrections, and we have developed a systematic approach to compute it for arbitrary values of the coupling. At weak coupling, our expansion matches perfectly with the available field-theoretic results up to three-loop order. At strong coupling, we have presented concrete predictions that await independent verification. The scaling function $d(g)$ thus furnishes a valuable example at the level of OPE coefficients that can be evaluated at any coupling and offers a nontrivial probe of the AdS/CFT correspondence.

Several interesting directions emerge from this work. A particularly pressing task is to perform an independent strong-coupling computation of the corresponding correlation function. The physical process involved is the emission of a GKP string from a D3-brane; it would be highly desirable to verify the scaling behavior and extract the associated exponent for direct comparison with our results.

It would also be important to gain a deeper physical understanding of the observed scaling behavior. In the case of anomalous dimensions, the cusp anomalous dimension admits an interpretation as the energy density of a flux tube. It would be intriguing to identify the analogous physical process underlying the scaling behavior of the OPE coefficient studied here. Along these lines, an interesting observation is that replacing maximal giant gravitons with non-maximal ones alters the scaling behavior significantly \cite{He:2026ios} and renders the problem more complex. Correspondingly, the associated boundary state is no longer integrable \cite{Chen:2019gsb}. We also note that three-point functions involving two single-trace BPS operators and one twist-two operator in the same limit exhibit a more intricate large spin structure \cite{Alday:2013cwa}. These observations single out the maximal giant graviton as a particularly simple and special case, whose internal structure appears to be remarkably tractable.

Beyond the leading term, it seems plausible that subleading corrections in the large-$S$ expansion may also be accessible. Such corrections would generally depend on the twist and would no longer be universal; nevertheless, an all-coupling computation of these subleading terms would be a valuable extension. 

The giant graviton one-point function, computed via the overlap of an integrable boundary state with an on-shell Bethe state, belongs to a broader class of one-point functions that have recently attracted considerable attention. This class includes, among others, defect one-point functions \cite{deLeeuw:2015hxa,Buhl-Mortensen:2016pxs,Buhl-Mortensen:2017ind,Komatsu:2020sup}, 't Hooft loop one-point functions \cite{Kristjansen:2023ysz,Kristjansen:2024map}, and one-point functions on the Coulomb branch \cite{Ivanovskiy:2024vel,Coronado:2025xwk}. Some of these quantities have also been extended to the ABJM theory \cite{Yang:2021hrl,Yang:2021kot,Yang:2022dlk,Chen:2019kgc,Jiang:2023cdm,Zhang:2025yex,Kristjansen:2021abc,Linardopoulos:2022wol}. The techniques developed in the present work are expected to be generalizable to these related setups. It would therefore be interesting to investigate their large spin behavior and to determine whether analogous scaling laws emerge, with exponents that can be computed non-perturbatively. Such quantities are particularly attractive in this context: if, as we have conjectured here, they are free from finite-size corrections, then asymptotic results alone would suffice to make all-loop predictions, greatly simplifying the analysis.

\vspace{0.5cm} 
\noindent{\bf Acknowledgements.} We thank Yu Hao for initial collaboration on project. We thank Shota Komatsu and Didina Serban for helpful discussions. This is supported by the National Natural Science Foundation of China through Grant No. 12575073 and 12247103.

%\nocite{PalettaDuhPozsgayZadnik2025}
\bibliography{reference} 
\bibliographystyle{utphys}

\clearpage
\onecolumngrid
\setcounter{figure}{0}
\setcounter{table}{0}
\setcounter{equation}{0}
\renewcommand{\thefigure}{S\arabic{figure}}
\renewcommand{\thetable}{S\arabic{table}}
\renewcommand{\theequation}{S\arabic{equation}}
\renewcommand{\theHfigure}{supp.\arabic{figure}}
\renewcommand{\theHtable}{supp.\arabic{table}}
\renewcommand{\theHequation}{supp.\arabic{equation}}

\begin{center}
{\large\bf Supplemental Material for ``A universal scaling function for giant graviton OPE coefficient''}
\end{center}
\section{Ratio of Gaudin-like determinants}
\begin{figure} 
    \centering       
    \includegraphics[width=0.95\textwidth]{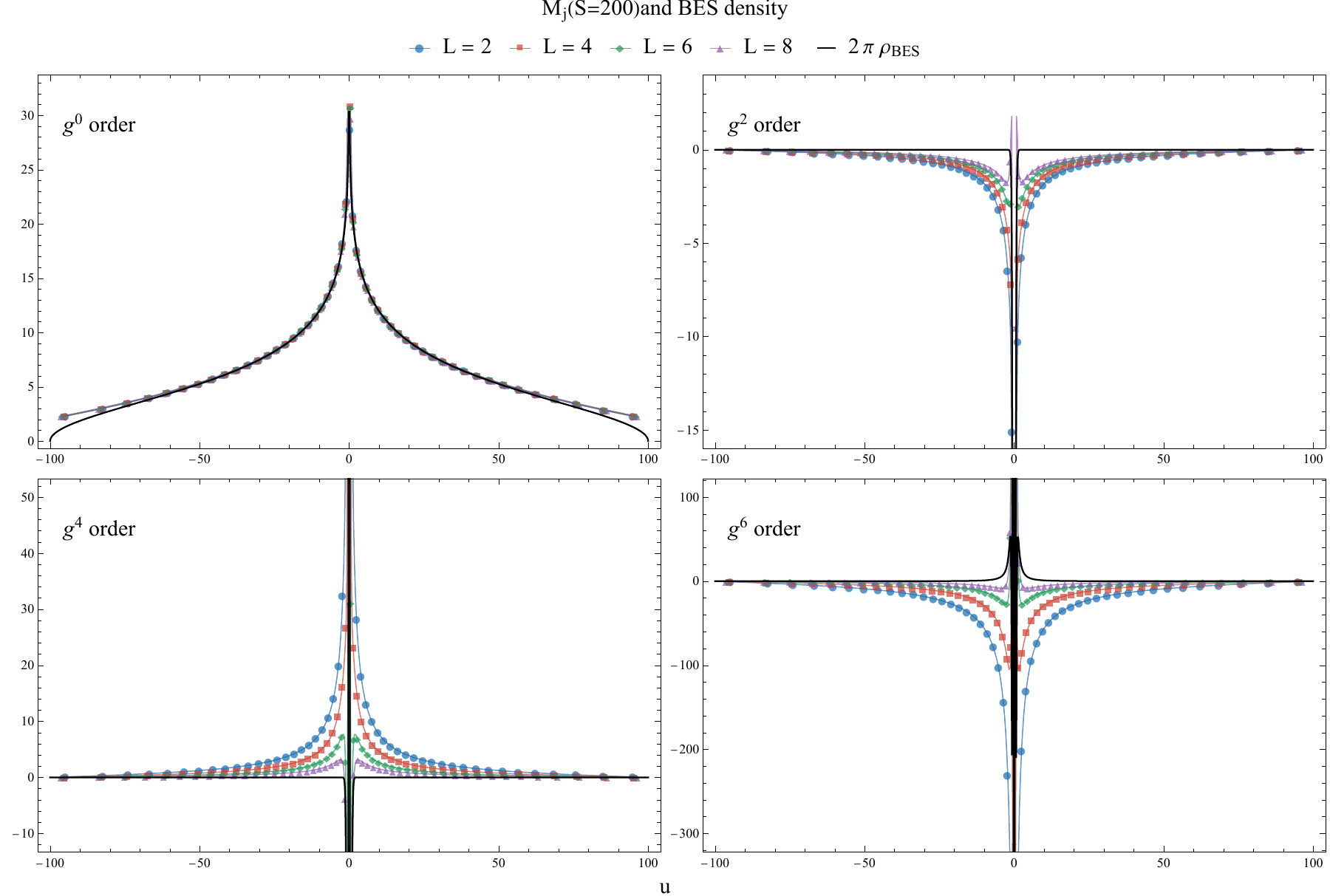}
    \caption{We plot $M_j$ for different twist $L=2,4,6,8$ with $S=200$ order by order. The results show that, up to order $g^6$, $M_j$becomes universal across different twists when $S$ is sufficiently large, and all cases converge to the BES density }
    \label{order_density_m} 
\end{figure}

The all-loop Gaudin-like matrices are
\begin{align}
\label{eq:supp_gaudin_det}
(G_{\pm})_{jk}
=M_j\delta_{jk}
-\left(\mathcal{K}_g(u_j,u_k)\pm \mathcal{K}_g(u_j,-u_k)\right),
\end{align}
where the scattering kernel is given by the derivative of the logarithmic all-loop S-matrix with dressing phase~\cite{Beisert:2006ez}
\begin{align}
\mathcal{K}_g(u,v)=-\ri \partial_{u}\log\left(\frac{x^{-}(u)-x^{+}(v)} {x^{+}(u)-x^{-}(v)} 
\frac{1-1/x^+(u)x^-(v)}
     {1-1/x^-(u)x^+(v)}e^{2\ri \theta(u,v)}
\right).
\end{align}
The determinants of all-loop Gaudin-like metrics can be written as 
\begin{align}
\det G_{\pm}=\prod_{j=1}^{S/2}M_j\,\times\det(1- K_{\pm}),\quad 
\left( K_{\pm}\right)_{jk}=\frac{1}{M_j}\left(\mathcal{K}_g(u_j,u_k)\pm \mathcal{K}_g(u_j,-u_k)\right). 
\end{align}
After factoring out the diagonal weights $M_j$, the ratio of Gaudin-like determinants can be expressed in terms of the scattering kernel
\begin{align}
\label{eq:supp_trace}
\mathcal{R}(S)=\frac{1}{2}\log\left(\frac{\det G_+}{\det G_-}\right)=\sum_{n=1}^\infty\frac{(\text{Tr}(K^+)^n-\text{Tr}(K^-)^n)}{2n}.
\end{align} 
By employing the BAEs, the diagonal component $M_j$, defined as the derivative of the logarithmic asymptotic BAEs, can be expressed as
\begin{align}
M_j=2\pi\frac{\rd\tilde{n}_k}{\rd u_k},
\end{align}
where $\tilde{n}_k$ are the mode numbers
\begin{align}
    \tilde{n}_k=k'+\frac{L-2}{2}\text{sign}(k'),\quad k'=\pm1\pm2,...,\pm\frac{S}{2}.
\end{align}
By definition, $M_j$ is related to the density in the large spin limit
\begin{align}
\lim_{S\to\infty}M_j=2\pi\lim_{S\to\infty} \frac{\rd \tilde{n}_k}{\rd u_k}
=2\pi\left(\rho(u_j)+\rho_{\text{hole}}(u_j)\right).
\end{align}
For the ground state with finite $L$, the hole density is vanishing and therefore does not contribute to the bulk distribution, so that we have
\begin{align}
\lim_{S\to\infty}M_j=2\pi\rho(u_j).
\end{align}
This relation is also checked numerically. We plot $M_j$ and BES density order by order in~\ref{order_density_m}. The result shows $M_j$ match the BES density order by order and become $L$-independent.

In the large spin limit, the trace terms in~\eqref{eq:supp_trace} become integral against the density. The remaining factor $1/M_j$ in $K_{\pm }$ cancels the density in the integral measure $\rho(u)\rd u$. This is the same as that in the tree-level case. The ratio of determinants is controlled by the normalized scattering kernel itself, not by the detailed density profile and the Bethe roots information. Finally, the all-loop ratio of Gaudin-like determinants can be expressed as
\begin{align}
\mathcal{R}(S)=-\sum_{n=1}^{\infty}\frac{1}{2 n}\int_{-S/2}^{S/2}\frac{\rd u_1}{2\pi}\int_{-\infty}^{\infty}\frac{\rd u_2}{2\pi}\cdots\frac{\rd u_n}{2\pi}\Big(\mathcal{K}_{g}(u_1,u_2)\mathcal{K}_{g}(u_2,u_3)\dots\mathcal{K}_{g}(u_n,-u_1)\Big),
\end{align}
where we have extended the integration range of $u_2,u_3,...,u_n$ to $(-\infty,\infty)$, while keeping that of $u_1$ as $(-S/2,S/2)$ to capture the large-$S$ behaviour.

For the $g^{0}$ order, the contribution arises solely from the kernel $\varphi(u-v)$. The ratio of determinants can be computed as following
\begin{align}
\mathcal{R}_0(S)=-\sum_{n=1}^{\infty}\frac{1}{2n}\int_{-S/2}^{S/2}\frac{\rd u_1}{2\pi}\int_{-\infty}^{\infty}\frac{\rd u_2}{2\pi}\cdots\frac{\rd u_n}{2\pi}\Big(\varphi(u_1-u_2)\varphi(u_2-u_3)\dots\varphi(u_n+u_1)\Big),
\end{align}
which leads to the result in~\eqref{eq:tree_R}.
\par
The higher order contribution can be obtained by doing the expansion
\begin{align}
\mathcal{K}_{g}(u,v)=
\varphi(u-v)+\sum_{k=1}^{\infty}g^{2k}\mathcal{K}^{(2k)}(u,v).
\end{align}
The first two orders read
\begin{align}
\mathcal{K}^{(2)}(u,v)=&\frac{16\left[-1+4u(u-2v)\right]}{(1+4u^2)^2(1+4v^2)},\quad 
\mathcal{K}^{(4)}(u,v)=\frac{32\,F_4(u,v)}{(1+4u^2)^4(1+4v^2)^3},\nonumber\\
F_4(u,v)=&3-60u^2+80u^4-64u^6+48uv-64u^3v+512u^5v\nonumber\\
&+12v^2-624u^2v^2+576u^4v^2+768u^6v^2\nonumber\\
&+256uv^3-2048u^3v^3+64v^4-1280u^2v^4\nonumber\\
&+768uv^5-3072u^3v^5 .
\end{align}
We find that the $g^{2}$ order contribution comes from the one-insertion case
\begin{align}
\mathcal R^{(2)}(S)=&-\frac{1}{2}\sum_{m=0}^{\infty}\int_{-S/2}^{S/2}\frac{\rd u_1}{2\pi}\prod_{k=0}^{m+1}\frac{\rd u_m}{2\pi}\Big(\varphi(u_1,u_2)\varphi(u_2,u_3)...\varphi(u_m,u_{m+1})\mathcal{K}^{(2)}(u_{m+1},-u_1)\Big).
\end{align}
The $g^{4}$ order contribution contains two parts: the one-insertion of $\mathcal{K}^{(4)}$ and two-insertion of $\mathcal{K}^{(2)}$
\begin{align}
\mathcal R^{(4)}(S)
=&-\frac{1}{2}\sum_{m=0}^{\infty}\int_{-S/2}^{S/2}\frac{\rd u_1}{2\pi}\int\prod_{k=2}^{m+1}\frac{\rd u_k}{2\pi}\Big(\varphi(u_1,u_2)\varphi(u_2,u_3)...\varphi(u_m,u_{m+1})\mathcal{K}^{(4)}(u_{m+1},-u_1)\Big)\nonumber\\
&-\frac14\sum_{m_1,m_2=0}^{\infty}\int_{-S/2}^{S/2}\frac{\rd u_1}{2\pi}\int\prod_{k=2}^{m_1+m_2+1}\frac{\rd u_k}{2\pi}\Big(\varphi(u_1,u_2)...\varphi(u_{m_1-1},u_{m_1})\mathcal{K}^{(2)}(u_{m_1},u_{m_1+1})\nonumber\\
&\qquad \qquad \qquad \qquad\times \ \ \varphi(u_{m_1+1},u_{m_1+2})...\varphi(u_{m_1+m_2},u_{m_1+m_2+1})\mathcal{K}^{(2)}(u_{m_1+m_2+1},-u_{1})\Big).
\end{align}
Substituting the specific form of $\mathcal{K}^{(2)}$ and $\mathcal{K}^{(4)}$, we work out 
\begin{align}
\mathcal R^{(2)}(S)=\frac{2}{\pi S}+\mathcal{O}\left(S^{-2}\right),\qquad \mathcal R^{(4)}(S)=-\frac{\pi}{6 S}+\mathcal{O}\left(S^{-2}\right).
\end{align}
This calculation can be systematized at higher loop orders and in all the cases we find that the higher order contribution starts at order $1/S$. This gives strong support that the higher-loop orders have no $\log S$ contributions. 

\section{Prefactor}
\label{sec:supp-density-prefactor}
Including the all-order bulk dressing phase, the density fluctuation satisfies the BES equation~\cite{Beisert:2006ez}
\begin{align}
\hat\sigma(t)=\frac{t}{e^t-1}\left[K(2gt,0)
-4g^2\int_0^\infty\rd t'\,
K(2gt,2gt')\hat\sigma(t')\right],
\label{eq:supp-bes-equation}
\end{align}
where $\hat\sigma(t)$ is the Fourier-type transformed density fluctuation
\begin{align}
\hat{\sigma}(t)=e^{-t/2}\int_{-\infty}^{\infty}e^{-\ri t u}\sigma(u).
\end{align}
The BES equation can also be expressed as
\begin{align}
\label{eq:bes_numeric_s}
s(t) = K(2gt,0) - 4g^2\int_0^\infty dt'
K(2gt,2gt')\frac{t'}{e^{t'}-1}s(t'),\quad s(t) = \frac{e^t-1}{t}\,\hat{\sigma}(t).
\end{align}
The BES equation can be solved for both weak and strong coupling expansion~\cite{Beisert:2006ez,Basso:2007wd}. 

The boundary dressing phase can be written as
\begin{align}
\log\sigma_{B}(u)=-4\log 2E(u)-\log\sigma_B^{\rm{min}}(u),
\end{align}
where
\begin{align}
\log\sigma_B^{\rm{min}}(u)=\sum_{r=1}^{\infty}\ri (2r+1)c_{2r+1}(g)q_{2r+2}(u),\quad c_{2r+1}(g)=2\ri(-1)^r\int_0^\infty\frac{\rd t}{t}
\frac{J_{2r+1}(2gt)}{1+e^{t/2}}.
\end{align}
In terms of the Fourier transformation, the all-loop prefactor can be expressed as
\begin{align}
\mathcal{P}(S)=&\,\mathcal{P}_0(S)-2g^2E_0\int_0^\infty\rd t\,\frac{s(t)}{e^{t/2}+1}\rd u-4\log 2 f(g)\log S+\frac{S}{4}\int_{-\infty}^{\infty}\log\sigma_B^{\rm{min}}(u)\,\rho(u)\rd u\,.
\label{supp-prefactor}
\end{align}
where $f(g)$ is the usual large-spin scaling function of the anomalous dimension. For completeness, the weak-coupling calculation proceeds as follows.  We first expand the complete BES kernel in powers of $g$ and solve the equation recursively. In parallel, we expand the minimal boundary dressing phase to the same order. These two series are then substituted into Eqs.~\eqref{supp-prefactor}. After expanding the Bessel functions, the remaining integrals can be evaluated algebraically.  A similar expansion in powers of $1/g$ can be performed in the strong-coupling regime. We carry out this procedure through $g^{14}$, i.e.\ seven
loops. Since the ratio of determinant contribution is shown to remain $-\frac14\log S$, the result for OPE scaling function is below
\begin{align}
d(g)|_{0<g\ll1}&=\frac14-4\log2\,g^2
+\frac43\left(\pi^2\log2-9\zeta(3)\right)g^4
+\frac4{45}\left(-11\pi^4\log2+30\pi^2\zeta(3)+1350\zeta(5)\right)g^6\\
&+\frac4{315}\Bigl(73\pi^6\log2-119\pi^4\zeta(3)-2100\pi^2\zeta(5)
+315\left(8\log2\,\zeta(3)^2-315\zeta(7)\right)\Bigr)g^8\nonumber\\
&+\Bigl(-\frac{14192}{14175}\pi^8\log2
+\frac{1216}{945}\pi^6\zeta(3)
+\frac{712}{45}\pi^4\zeta(5)+96\zeta(3)^3\nonumber\\
&\quad+\pi^2\left(280\zeta(7)-\frac{64}{3}\log2\,\zeta(3)^2\right)
-640\log2\,\zeta(3)\zeta(5)
+14280\zeta(9)\Bigr)g^{10}\nonumber\\
&+\Bigl(\frac{547532}{467775}\pi^{10}\log2
-\frac{19324}{14175}\pi^8\zeta(3)
-\frac{13144}{945}\pi^6\zeta(5)
+\frac8{15}\pi^4\left(32\log2\,\zeta(3)^2
-315\zeta(7)\right)\nonumber\\
&\quad-\frac83\pi^2\left(
20\zeta(3)\left(\zeta(3)^2-8\log2\,\zeta(5)\right)
+1197\zeta(9)\right)\nonumber\\
&\quad+192\left(-16\zeta(3)^2\zeta(5)
+17\log2\,\zeta(5)^2
+35\log2\,\zeta(3)\zeta(7)\right)-171864\zeta(11)\Bigr)g^{12}
\nonumber\\
&+\Bigl(-\frac{61440712}{42567525}\pi^{12}\log2
+\frac{759376}{467775}\pi^{10}\zeta(3)
+\frac{212936}{14175}\pi^8\zeta(5)
-\frac{32}{189}\pi^6\left(94\log2\,\zeta(3)^2-871\zeta(7)\right)
\nonumber\\
&\quad+\frac8{15}\pi^4\left(
68\zeta(3)^3-656\log2\,\zeta(3)\zeta(5)
+3619\zeta(9)\right)+8\pi^2\Bigl(208\zeta(3)^2\zeta(5)
-272\log2\,\zeta(5)^2\Bigr)\nonumber\\
&\quad+8\pi^2\left(-560\log2\,\zeta(3)\zeta(7)+4851\zeta(11)\right)
-256\log2\left(
\zeta(3)^4+273\zeta(5)\zeta(7)
+294\zeta(3)\zeta(9)\right)\nonumber\\
&\quad+48\left(
704\zeta(3)\zeta(5)^2+700\zeta(3)^2\zeta(7)
+45045\zeta(13)\right)\Bigr)g^{14}+\mathcal O(g^{16}).\nonumber
\end{align}
\section{Numerical result for finite coupling}
The finite-coupling computation follows the Bessel-matrix method proposed in~\cite{Benna:2006nd}.  Define
\begin{align}
Z_{mn}(g)&=\int_0^\infty\frac{J_m(2g t)J_n(2g t)}{t(e^t-1)}\rd t,\quad s(t)=\sum_{n=1}^{\infty}s_n\frac{J_n(2g t)}{2g t}.
\end{align}
Let $N={\rm diag}(1,2,3,\ldots)$, $P$ project onto even Bessel modes,
and $Q$ project onto odd Bessel modes.  With
\begin{align}
C=2PNZQN,
\end{align}
where
\begin{align}
P=\text{diag}(0,1,0,1,...),\quad Q=\text{diag}(1,0,1,0,...),\quad N=\text{diag}(1,2,3,...)\,.
\end{align}
After projecting the BES equation~\eqref{eq:bes_numeric_s} on Bessel modes, it becomes a matrix equation, whose solution can be written as
\begin{align}
\mathbf{s}=\left(1+2NZ+4CZ\right)^{-1}(1+2C)e,\quad e=(1,0,0,\ldots)^T \,.
\end{align}
The all-order cusp anomalous dimension is
\begin{align}
f(g)=8g^2s_1.
\end{align}
With this density representation, we can express the scaling function in terms of Bessel modes
\begin{align}
d(g)=\frac{1}{4}
-4g^2\,\mathbf s^T\mathbf h+4g\left(a_1(g)-\mathbf s^T Z\mathbf b\right)
-8\left(\log 2\right)g^2s_1,
\end{align}
where
\begin{align}
h_n=\int_0^\infty\frac{J_n(2g t)}{2g t\,(e^{t/2}+1)}\,dt,\quad a_n(g)=\int_0^\infty\frac{dt}{t}\frac{J_n(2gt)}{1+e^{t/2}},\quad 
b_n=
\begin{cases}
2n\,a_n(g)& n\ {\rm odd},\\
0& n\ {\rm even}.
\end{cases}
\end{align}
Then we can numerically compute the full prefactor using the Bessel basis projection. In~\cite{Benna:2006nd}, it shows that the matrix elements of $Z$ decay sufficiently fast with increasing dimension. We truncate the series expansions of each term and of the kernel after the first 80 orders of Bessel functions, which means that the dimension of the matrices and the vectors is 80. The numerical result for finite $g$ is shown in Figure~\ref{all_order_scaling_function}.

\end{document}